\documentclass[aps,pra,floatfix,twocolumn,superscriptaddress]{revtex4}

\usepackage{graphicx}
\usepackage{graphics}
\usepackage{amssymb}
\usepackage{amsmath} 
\usepackage{epsfig}
\usepackage{latexsym}
\usepackage{color}
\usepackage{subfigure}
\usepackage{array}

\newcommand{\ket}[1]{|#1\rangle}

\newcommand{\expct}[1]{\langle #1 \rangle}
\newcommand{\Tr}[1]{\mathrm{Tr}\left( #1 \right)}

\begin{document}

\title{Gates for the Kane Quantum Computer in the Presence of Dephasing}

\author{Charles D. Hill} 
\email{hillcd@physics.uq.edu.au}
\affiliation{Centre for Quantum Computer
Technology, and Department of Physics, The University of Queensland, St
Lucia, QLD 4072, Australia} 

\author{Hsi-Sheng Goan}
\email{goan@physics.uq.edu.au}
\affiliation{Centre for Quantum Computer Technology, University of
New South Wales, Sydney, NSW 2052, Australia}
\thanks{Mailing Address: Centre for Quantum Computer
Technology, C/- Department of Physics, The University of Queensland, St
Lucia, QLD 4072, Australia}


\begin{abstract}
In this paper we investigate the effect of dephasing on proposed
quantum gates for the solid-state Kane quantum computing
architecture. Using a simple model of the decoherence, we find that
the typical error in a CNOT gate is $8.3 \times 10^{-5}$. We also
compute the fidelities of Z, X, Swap, and Controlled Z operations
under a variety of dephasing rates. We show that these numerical
results are comparable with the error threshold required for fault
tolerant quantum computation.
\end{abstract}

\maketitle


\section{Introduction}

One of the most exciting advances in physics has been the development
of quantum algorithms \cite{Sho97, Gro97} which outperform their best
known classical counterparts. These algorithms are described in the
absence of noise, and decoherence. In experiment this will certainly
not be the case. In this paper we investigate numerically how a simple
model of decoherence affect gates on the Kane quantum computer
\cite{Kan98}. The Kane quantum computer is one of a number of
promising silicon based quantum computer proposals \cite{SDK03, LGY+02,
DSD03, VRYE+00, FRS+03}.

Dephasing in systems similar to the Kane architecture has been
investigated since the introduction of the spin echo technique
\cite{Hah50}. The nuclear spin and electronic spin decoherence times
of $\mathrm{P}$ donors in $\mathrm{Si}$ is relatively long
\cite{Hon54, GB58, FG59, Feh59, HSE60, Fau69, CH72, WJS88}. For
example, recently a time of $60\mathrm{ms}$ was measured for the
electronic dephasing time, $T_{2_e}$ \cite{TLA+03}. Although dephasing
times are comparatively long, if left unchecked the accumulated errors
introduced by dephasing will destroy coherence in the computation.

Using quantum error correction protocols it may be possible to correct
the errors caused by decoherence. \cite{Sho95, Ste96a, CS96, Ste96b,
KLM+01}. To successfully reduce the overall error in the system, we
must correct errors faster than they accumulate, in a fault-tolerant
manner \cite{Sho96}. This consideration leads to an error threshold
\cite{AB97, Got97, Pre99, KLZ98}. Typically such
a threshold requires the probability of introducing an error in each
gate to be below $1\times 10^{-4}$ to as low as $1 \times
10^{-6}$. For the Kane architecture the exact threshold is still under
investigation \cite{Fow03}. In this paper we ask, if it is possible
for the Kane architecture to achieve this error threshold. To do this
we must know how much error is introduced by each of the gates used in
the Kane quantum computing architecture.

This paper shows the results of simulations of gates on the Kane
architecture in the presence of dephasing. Single qubit gates
presented here are similar to those used in Nuclear Magnetic Resonance
(NMR) \cite{CLK+01, Sli78} for rotations of individual qubits.
Voltage fluctuations on the `A' gate and stochastic modeling of the
system has been investigated analytically in Refs. \cite{WH02,
Wel01}. The two qubit gates presented here use non-adiabatic pulse
schemes \cite{HG03}. 

The analysis given here is a direct analogue to that of Fowler et al
\cite{FWH03} for the adiabatic CNOT gate. The gates analyzed in this
paper are simpler, faster, and potentially higher fidelity than the
adiabatic gate \cite{HG03}. These gates do not rely on complicated
pulse shapes, but simply turning on or off the voltages applied. In
contrast to adiabatic gates, the timing of these gates could easily be
run off a digital clock cycle. In addition, whereas the adiabatic CNOT
is required to be applied up to three times to create an arbitrary two
qubit gate, using non-adiabatic schemes it is possible to create an
arbitrary two qubit gate directly. For example, the swap gate analyzed
here would require three adiabatic CNOT gates to construct. Using
non-adiabatic schemes we are able to construct it in a single pulse
sequence \cite{HG03}, which is much faster, simpler and higher
fidelity than the corresponding adiabatic scheme.

We compare each of the gates analyzed to the error threshold for fault
tolerant quantum computing. We simulate the master equation for
typical values of spin dephasing expected in the Kane architecture. We
find that the error in the gates analyzed is less than or comparable
to that required for fault tolerant quantum computation.

This paper is organized as follows. In Section
\ref{sec:MasterEquation} we describe the simple model of decoherence
which we use and present the master equation for the system. Section
\ref{sec:OneQubit} presents the results for one qubit gates, including
free evolution in subsection \ref{ssec:FreeEvolution}, Z rotations in
subsection \ref{ssec:ZRot} and X rotations in subsection
\ref{ssec:XRot}. Two qubit non-adiabatic gates are shown in Section
\ref{sec:TwoQubit}. These include the CNOT gate in subsection
\ref{ssec:CNOT}, and the swap gate and controlled Z gates in
subsection \ref{ssec:Swap}. Finally, conclusions are drawn in Section
\ref{sec:Conclusion}.


\section{The Master Equation} \label{sec:MasterEquation} 

A brief introduction to the Kane quantum computing architecture
\cite{Kan98} is given here. The Kane architecture consists of
$\mathrm{P}$ donor atoms embedded in $\mathrm{Si}$. The orientation of
the nuclear spin of each $\mathrm{P}$ donor represents one qubit.

%
%
When placed in a magnetic field applied in the $z$ direction, Zeeman
splitting occurs. This is given by the Hamiltonian
\begin{equation}
H_B = - g_n \mu_n B Z_n + \mu_B B Z_e,
\end{equation}
where $Z$ is the Pauli Z matrix, and the subscripts $e$ ($n$) indicate
electronic (nuclear) spin. The magnetic field, $B$, may be controlled
externally. The application of a resonant rotating magnetic field adds
the following terms to the spin Hamiltonian:
\begin{eqnarray}
H_{ac} &= - g_n \mu_n B_{ac} &\left[ X_n \cos(\omega_{ac}t) 
	+ Y_n \sin(\omega_{ac}t) \right] \nonumber\\
      & + \mu_B B_{ac} & \left[X_e \cos(\omega_{ac}t) + Y_e
	\sin(\omega_{ac}t) \right].
\end{eqnarray}
The electronic spins couple to their corresponding nuclear spins via
the hyperfine interaction,
\begin{equation}
H_A = A \mathbf{\sigma_e} \cdot \mathbf{\sigma_n},
\end{equation}
where $A$ is the strength of the interaction. The design of the Kane
quantum computer calls for control of the strength of the hyperfine
interaction externally by applying appropriate voltages to `A'
gates. The electronic spins couple to adjacent electrons via the
exchange interaction
\begin{equation}
H_J = J \mathbf{\sigma_{e_1}} \cdot \mathbf{\sigma_{e_2}},
\end{equation}
where $e_1$ and $e_2$ are two adjacent electrons, and J is the
strength of the exchange interaction which may be controlled
externally through the application of voltages to the `J' gates.

Altogether the spin Hamiltonian of a two donor system is given by 
\begin{equation}
H_s = \sum_{i=1}^2 H_{B_i} + H_{A_i} + H_{J} + H_{ac_i}. 
	\label{eqn:sysHamiltonian}
\end{equation}

The times and fidelities of the gates naturally depend on exactly
which parameters are used to calculate them. For many of the gates in
this paper the typical parameters shown in Table \ref{tab:Parameters}
were used. These parameters are similar to the parameters used for the
pure state calculations in \cite{HG03}.

\begin{table}
\begin{tabular}{|>{\centering}m{4cm}|>{\centering}m{1cm}|c|}
\hline \textbf{Description} & \textbf{Term} & \textbf{Value} \\
\hline
Unperturbed Hyperfine Interaction & $A$ & $0.1211 \times
10^{-3} \textrm{meV}$ \\ 
\hline 
Hyperfine Interaction During Z Rotation & $A_z$ & $0.0606 \times
10^{-3} \textrm{meV}$\\ 
\hline 
Hyperfine Interaction during X Rotation & $A_x$ & $0.0606 \times
10^{-3} \textrm{meV}$\\ 
\hline
Constant Magnetic Field Strength & $B$ & $2.000 \textrm{T}$ \\
\hline
Rotating Magnetic Field Strength & $B_{ac}$ & $0.0025 \textrm{T}$ \\ 
\hline 
Hyperfine Interaction during Interaction  & $A_U$ &  $0.1197 \times
10^{-3} \  \textrm{meV}$ \\
\hline
Exchange Interaction during Interaction & $J_U$ & $0.0423 \  \textrm{meV}$\\
\hline
\end{tabular}
\caption{Typical parameters used for numerical calculations.}
\label{tab:Parameters}
\end{table}

%
%

A simple model of decoherence was used for these calculations. There
are many different decoherence mechanisms, but our model only
considers pure dephasing (without engergy relaxation). Whereas
dephasing is certainly not the only source of decoherence, it likely
to be the dominant effect on a time scale shorter than the energy
relaxation (dissipation) time, $T_1$. For example, Feher and Gere
\cite{FG59} measured $T_{1_{n}} > 10 \, \mathrm{hours}$ for nuclear
spin at a temperature of $T=1.25\mathrm{K}$, $B=3.2\mathrm{T}$ and
$T_{1_e}\approx 30 \, \mathrm{hours}$ under similar conditions. In
contrast, experimentally measured times for $T_2$ have been much
shorter. Gordon and Bowers \cite{GB58} measured $T_{2_e}=520 \mu s$
for $\mathrm{P:Si}$ at $T = 1.4\mathrm{K}$ in isotopically enriched
$\mathrm{^{28}\mathrm{Si}}$. Chiba and Harai \cite{CH72} have also
measured the electronic decoherence times of $\mathrm{P:Si}$, finding
a rate of $T_{2_e} = 100 \mu s$. For the nucleus, recent results for
the nuclear spin of a $\mathrm{^{29}Si}$ nucleus show a maximum value
of $T_{2_n}=25s$ \cite{LMA+03}.

Recently Tyryshkin et al \cite{TLA+03} obtained an experimental
measurement of $T_{2_e} = 14.2 \mathrm{ms}$ at $T=8.1K$ and
$T_{2_e}=62\mathrm{ms}$ at $T=6.9K$ for a donor concentration of $0.87
\times 10^{15} \mathrm{cm^{-3}}$ in isotopically pure
$\mathrm{Si}$. At millikelvin temperatures the decoherence time is
likely to be even longer. Additionally these measurements were carried
out in a bulk doped sample, and in our case we will be considering a
specifically engineered sample. Interactions such as exchange and
dipole-dipole interactions contribute to the coupling between
electrons. In some experiments, such as in Ref. \cite{TLA+03}, these
potentially beneficial coupling have been treated as sources of
decoherence, but in the operation of a quantum computer these
interactions can be either be decoupled or used to generate
entanglement, useful for quantum computation. Hence, it is expected
that $T_{2_e}$ may even be longer than those reported in
\cite{TLA+03}. Nevertheless, we use the value of $60 \mathrm{ms}$ as a
conservative estimate for electronic dephasing time. We expect the
nuclear dephasing times to be several orders of magnitude bigger than
electronic dephasing times. We choose the following parameters to be
typical of $\mathrm{P:Si}$ the systems we are considering:
\begin{eqnarray}
T_{2_{e}} &=& 60 \mathrm{ms}, \label{eqn:typicalDephasing1}\\
T_{2_{n}} &=& 1 \mathrm{s}. \label{eqn:typicalDephasing2}
\end{eqnarray}
The typical errors presented in the tables contained in the next three
sections are evaluated at these typical dephasing times.

The simple decoherence model we consider corresponds to the master
equation
\begin{equation}
\dot{\rho} = -\frac{i}{\hbar}[H_s, \rho] - \mathcal{L}[\rho],
\label{eqn:Master}
\end{equation}
where the dephasing terms are given by
\begin{eqnarray}
\mathcal{L}[\rho]&=&\sum_{i=1}^2 \Gamma_e \left[Z_{e_i},[Z_{e_i},\rho] \right] 
	+ \Gamma_n\left[Z_{n_i}, [Z_{n_i},\rho]\right] \label{eqn:L}
\end{eqnarray}
Characteristic dephasing rates, $\Gamma_{2_e}$ and $\Gamma_{2_n}$ are related to
the dephasing rates by the equations:
\begin{eqnarray}
T_{2_e} = \frac{1}{4 \Gamma_e},\\
T_{2_n} = \frac{1}{4 \Gamma_n}.
\end{eqnarray}

We define fidelity (and therefore error) in terms of the actual state
after applying an operation, $\rho$, and the intended state after that
operation, $\rho'$. Due to systematic errors and decoherence these
states will not necessarily be the same. When comparing against a pure
state $\rho'$, the fidelity $F$ of an operation is defined as
\begin{equation}
F(\rho, \rho') = \Tr{\rho \rho'};
\end{equation}
Error is defined in terms of fidelity
\begin{equation}
E(\rho, \rho') = 1 - F(\rho, \rho').
\end{equation}

Typically we would like to know the greatest error possible for any
input state. This is a computationally difficult problem. In the
results which follow, the approach taken is to calculate the fidelity
for each of the computational basis states, and each of the input
states which would ideally generate a Bell state. This has two main
benefits. The first is that a high fidelity indicates that the gate is
successfully creating or preserving entanglement. The second is that
Bell states are superposition states, which are susceptible to
dephasing. We also calculated the effect of each gate on the four Bell
input states for the CNOT gate. For typical parameters, Bell input
states give similar fidelities to those shown in this paper.

Throughout this paper we will use the states $\ket{0}$ and $\ket{1}$
to represent the nuclear spin up and spin down states respectively. We
will use the $\ket{\uparrow}$ and $\ket{\downarrow}$ to represent
electronic spin up and spin down states respectively.


\section{One Qubit Gates} \label{sec:OneQubit} 

\subsection{Free Evolution} \label{ssec:FreeEvolution} 

The spin of an isolated nucleus undergoing Larmor precession in the
presence of a magnetic field, $B$ \cite{GDJ95}. In this case it is
easy to solve the master equation with dephasing exactly. Considering
only the nuclear spin, we have
\begin{equation}
H_s = g_n \mu_n B Z_n.
\end{equation} 
The decoherence terms has only the single term
\begin{eqnarray}
\mathcal{L}[\rho] &=& \Gamma_n[Z_n, [Z_n, \rho]],\\
	&=& 2\Gamma_N (\rho - Z_n \rho Z_n).
\end{eqnarray}

In the rotating frame, the master equation has the solution
\begin{equation}
\rho(t) = \left[
	\begin{array}{cc}
	\rho_{00}(0)                  & \rho_{01}(0) e^{-4\Gamma_N t} \\
	\rho_{10}(0) e^{-4\Gamma_N t} & \rho_{11}(0)
	\end{array}
	\right].
\end{equation}
This has the effect of exponentially decaying the off diagonal terms
of the density matrix, but leaves the diagonal components
unchanged. For a single isolated nuclear spin, the simple model has no
effect on eigenstates of $Z_n$ (i.e., there is no relaxation process for
these states). In contrast, it has a dramatic influence on
superposition states whose off diagonal terms decay exponentially
(i.e., dephasing). Two such states are
\begin{eqnarray*}
\ket{+} = \frac{1}{\sqrt{2}} \left(\ket{0} + \ket{1}\right), \\
\ket{-} = \frac{1}{\sqrt{2}} \left(\ket{0} - \ket{1}\right).
\end{eqnarray*}
We can easily calculate the expectation value of the Pauli $X$ matrix,
$\expct{X}$, for the $\ket{+}$ state
\begin{eqnarray}
\expct{X} &=& \Tr{X \rho}, \\
&=& \exp(-4 \Gamma_N t).
\end{eqnarray}

\begin{figure}
\begin{center}
\includegraphics[width=8cm]{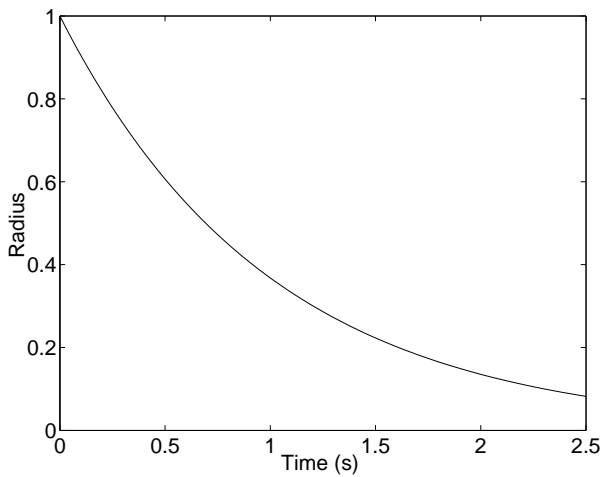}
\caption{Effect of dephasing on the free evolution of a single
embedded P atom in the Kane architecture.}
\label{fig:zTypical}
\end{center}
\end{figure}

For a single nuclear spin coupled to an electronic spin via the
hyperfine interaction the Hamiltonian is given by
\begin{equation}
H_s = H_B + H_A.
\end{equation}
We assume that electron is initially polarized by the large magnetic
field, $B$. The evolution of this Hamiltonian was calculated for their
typical values [given in Eq. (\ref{eqn:typicalDephasing1}),
Eq. (\ref{eqn:typicalDephasing2}) and Table \ref{tab:Parameters}]. The
fidelity after different times is shown in Fig.
\ref{fig:zTypical}. This figure shows the Bloch sphere radius, given
by
\begin{equation}
r = \sqrt{\expct{X}^{2} + \expct{Y}^{2} + \expct{Z}^{2}}.
\end{equation}
A pure state has a radius of one, and a radius of less than one
indicates a mixed state. An initial state of $\ket{\downarrow +}$ was
used. The radius decays at a rate governed by the nuclear decoherence
time, which in this case is $T_{2_n} = 1\mathrm{s}$. As is expected,
this decay is the same as the well known solution to the Bloch
equations.

\subsection{Z Rotations} \label{ssec:ZRot}

\begin{figure}
\begin{center}
\includegraphics[width=8cm]{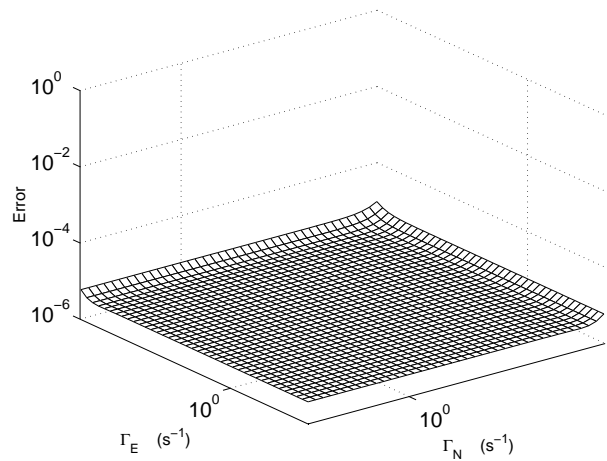}
\caption{Error in the Z gate for $\ket{\downarrow0}$ initial state at differing
rates of dephasing.}
\label{fig:z0}
\end{center}
\end{figure}

\begin{figure}
\begin{center}
\includegraphics[width=8cm]{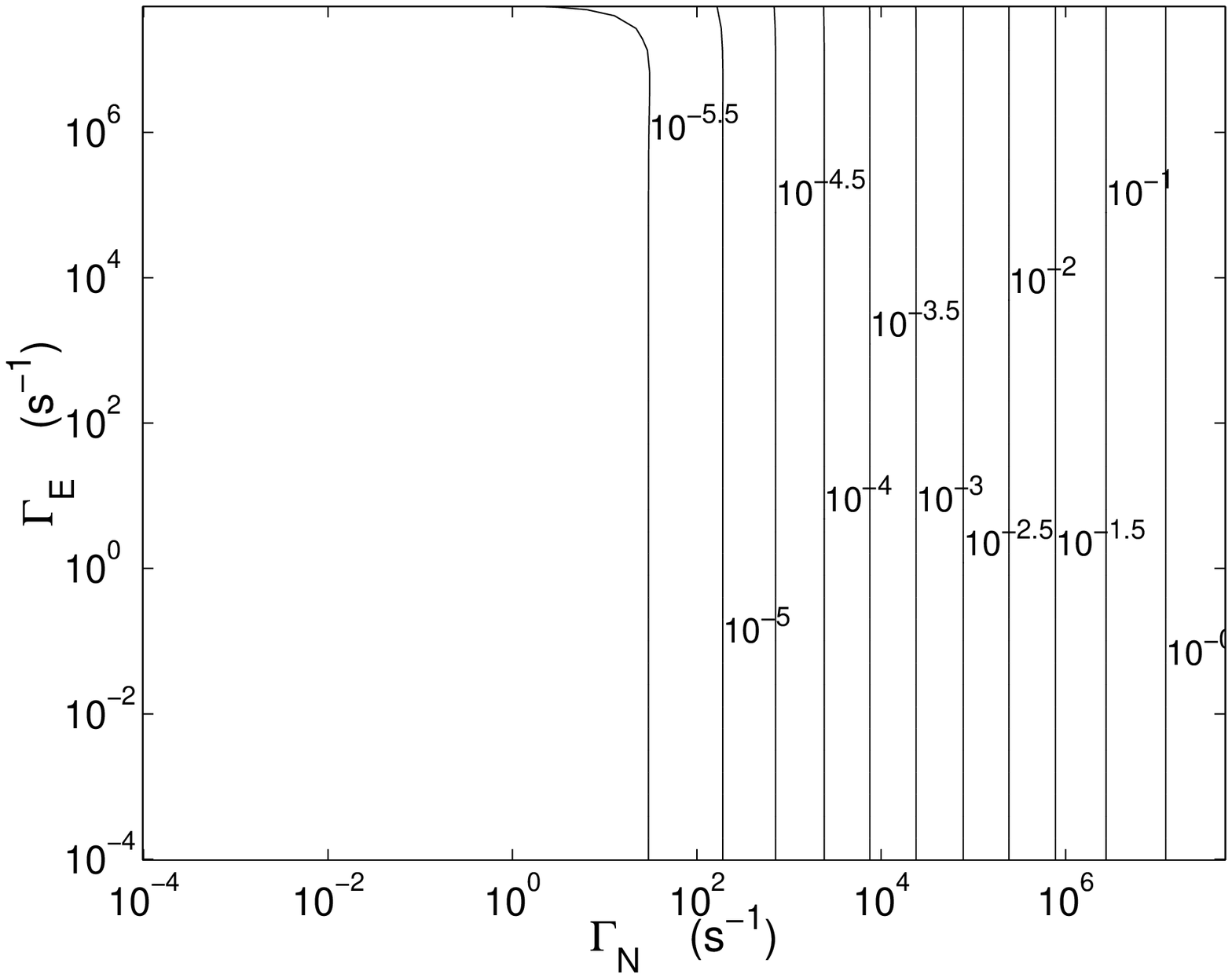}
\caption{Contour plot showing the error in the Z Gate for
$\ket{\downarrow+}$ initial state evolving at differing rates of
dephasing. Contour lines connect, and are labelled by, points of equal
error.}
\label{fig:zSym}
\end{center}
\end{figure}

Z rotations on the Kane architecture may be performed by varying the
Larmor precession frequency of a single qubit \cite{HG03}. Graphs
showing the error in the Z gate at different dephasing rates are
shown in Figs. \ref{fig:z0} and \ref{fig:zSym}.

Figure \ref{fig:z0} shows the error at different rates of decoherence
for both electrons, $\Gamma_e$, and nuclei, $\Gamma_n$, for a single
qubit in the $\ket{\downarrow0}$ state. The calculated error does not
significantly depend on the electronic or nuclear dephasing rate. The
pure $\ket{\downarrow0}$ state is not affected by decoherence terms in
the master equation. Therefore the only effect of dephasing occurs
when the hyperfine interaction rotates this state. The effect of
dephasing on this state is negligible.

The error in the Z operation for an initial state of
$\ket{\downarrow0}$, is primarily due to systematic error of $3.8
\times 10^{-6}$. This error is due to the hyperfine interaction
coupling between electrons and nuclei. This allows a small probability
of finding the electrons in an excited state. At typical rates of
decoherence for the short duration of a Z rotation (approximately
$21\mathrm{ns}$), systematic error is the dominant effect. For the
$\ket{\downarrow0}$ state, the typical error is $3.8 \times 10^{-6}$.

In the previous section we noted that superposition states,
$\ket{\downarrow+}$ and $\ket{\downarrow-}$, are affected by dephasing
terms more than eigenstates of $Z$. This is illustrated in
Fig. \ref{fig:zSym}. Electronic dephasing times have little effect on
the overall fidelity. As the nuclear dephasing rate increases, the
fidelity decreases, with a maximum error of approximately $0.5$. This
indicates all quantum coherence was lost and we are in classical
mixture of the states $\ket{\downarrow0}$ and $\ket{\downarrow1}$. For
typical rates of dephasing [given in
Eq. (\ref{eqn:typicalDephasing1}), and
Eq. (\ref{eqn:typicalDephasing2})], the error is found to be $1.9
\times 10^{-6}$ which is largely due to systematic error.

The maximum error, for typical rates of dephasing, of any of the
states tested for the Z gate is $3.8 \times 10^{-6}$. This error is
largely due to systematic effects rather than dephasing. This error
suggests it is theoretically possible to do a Z rotation with an error
of less than the $1 \times 10^{-4}$ limit suggested for fault tolerant
quantum computing. The results for the Z gate are summarized in Table
\ref{tab:Z}.

\begin{table}
\begin{tabular}{|c|c|c|}
\hline
State & Systematic Error & Typical Error \\
\hline
$\ket{0}$ & $3.8 \times 10^{-6}$ & $3.8 \times 10^{-6}$ \\
$\ket{+}$ & $1.9 \times 10^{-6}$ & $1.9 \times 10^{-6}$ \\
\hline
Maximum & $3.8 \times 10^{-6}$ & $3.8 \times 10^{-6}$\\
\hline
\end{tabular}
\caption{Summary of Z gate error.} \label{tab:Z}
\end{table}

\subsection{X Rotations} \label{ssec:XRot}

\begin{figure}
\begin{center}
\includegraphics[width=8cm]{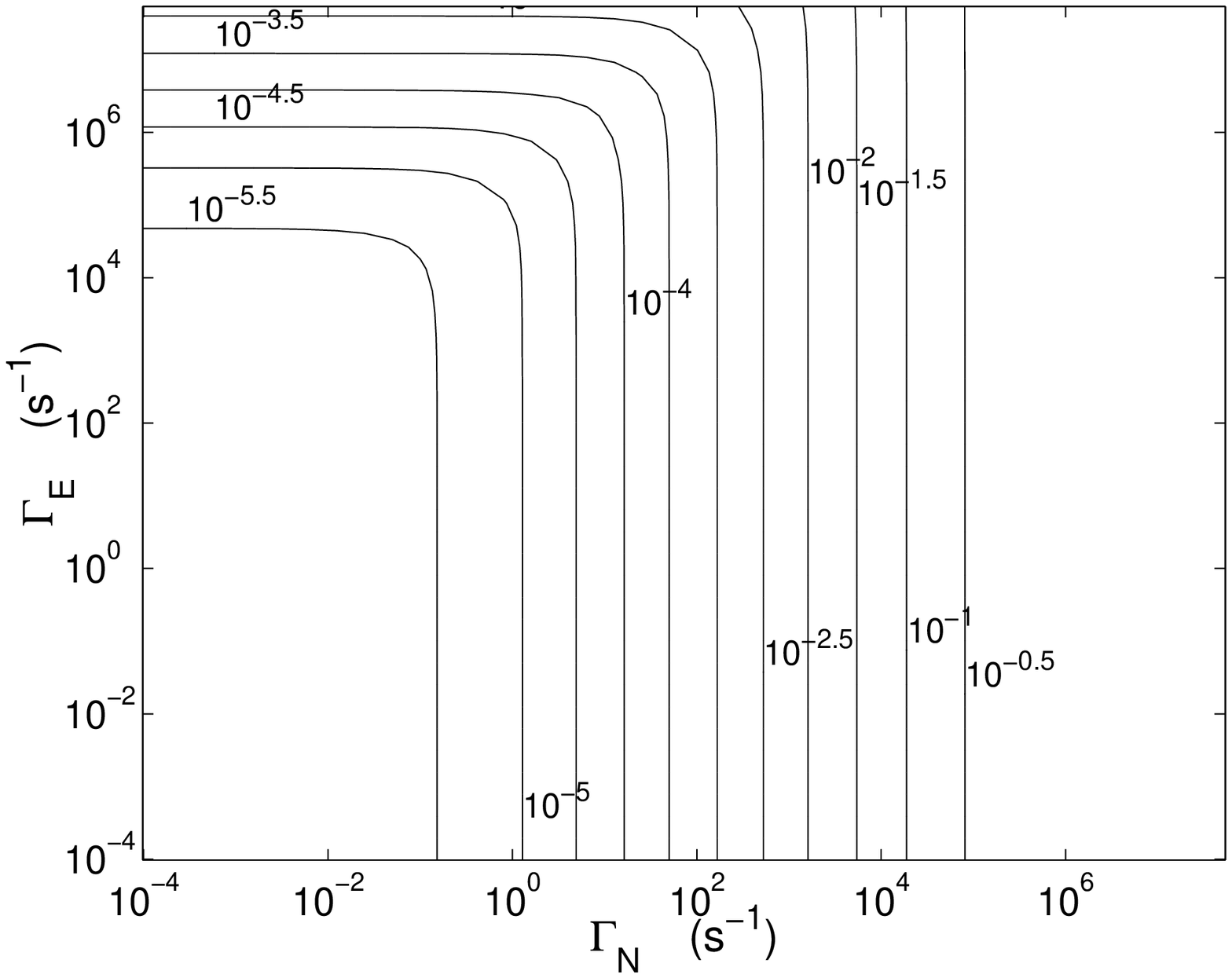}
\caption{Contour plot showing the error in the X gate for $\ket{\downarrow0}$
initial state evolving at differing rates of dephasing. Contour lines
connect, and are labelled by, points of equal error.}
\label{fig:x0}
\end{center}
\end{figure}

\begin{figure}
\begin{center}
\includegraphics[width=8cm]{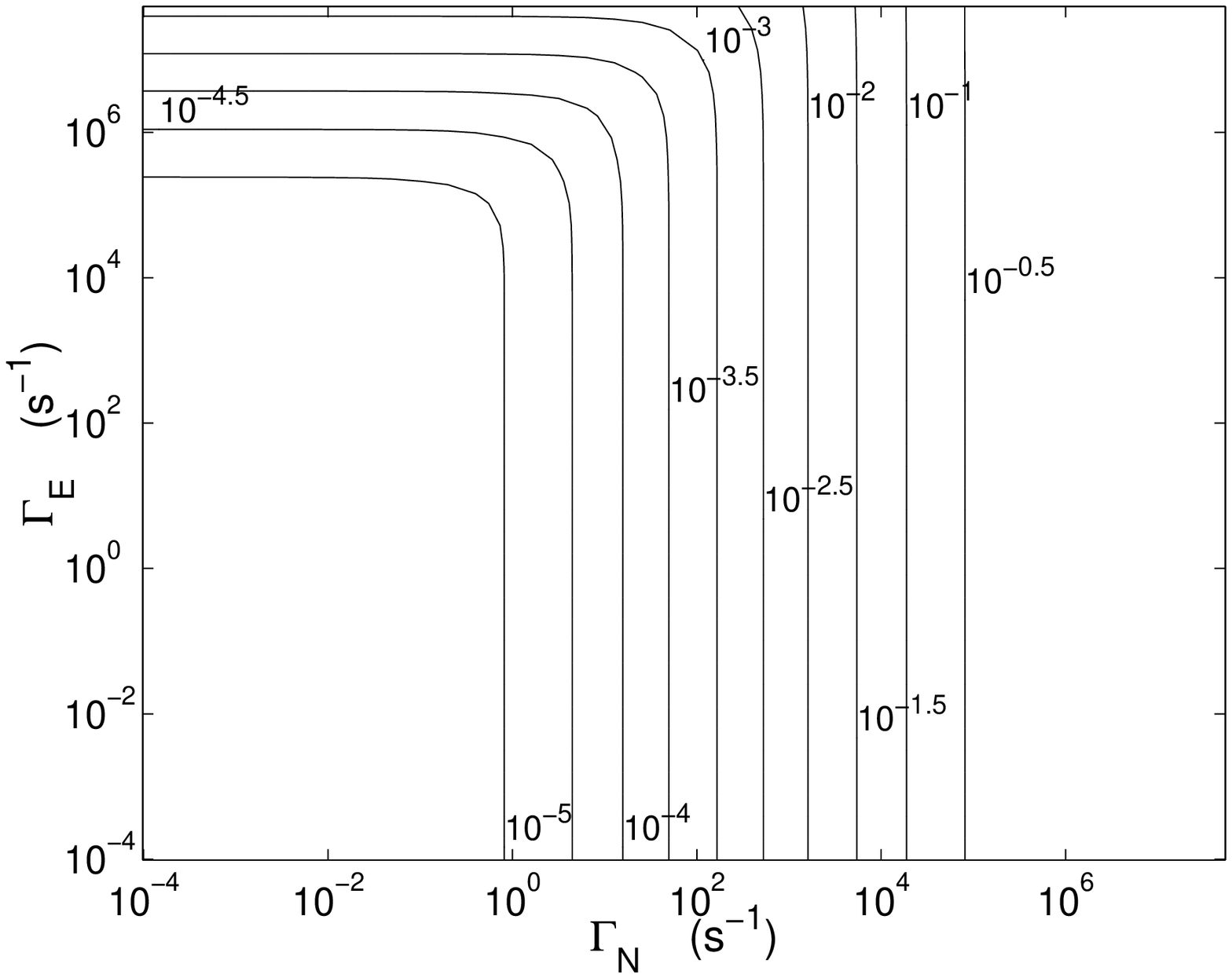}
\caption{Contour plot showing the error in the X Gate for $\ket{\downarrow1}$
initial state at differing rates of dephasing. Contour lines connect,
and are labelled by, points of equal error.}
\label{fig:x1}
\end{center}
\end{figure}

X rotations are performed using a resonant magnetic field,
$B_{ac}$. The error in the X rotation was found at different rates of
dephasing. This is shown for the two basis states in Figs.
\ref{fig:x0} and \ref{fig:x1}. Evident on both of these graphs is a
valley which levels out at a minimum error. This error is primarily
due to systematic error in the gate. For the $\ket{\downarrow0}$
initial state the systematic error is found to be $2.3 \times
10^{-6}$, and for the $\ket{\downarrow1}$ initial state the systematic error is
found to be $4.9 \times 10^{-6}$. As the nuclear dephasing rates,
$\Gamma_{2_n}$, increases, the fidelity of the operation
decreases. The fidelity of the operation drops to $0.0$ indicating
that, in the limit of large dephasing rates, the rotating magnetic field
does not have the desired effect of a resonant magnetic field.

In Figs. \ref{fig:x0} and \ref{fig:x1}, we see that the fidelity of
the X rotation depends weakly on the electronic dephasing rate,
$\Gamma_{2_e}$. The solutions to the equations under these conditions
show that the principal cause of error is the electron becoming
excited to a higher energy level.

Under typical conditions, the $\ket{\downarrow0}$ initial state has an
error of $3.8 \times 10^{-6}$. The $\ket{\downarrow1}$ state has an
error of $6.4\times 10^{-6}$. In this operation dephasing has a much
more important role than in Z rotations. One reason for this is that
an X rotation takes longer, approximately $6.4\mu s$.

The error in the X gate is summarized in Table \ref{tab:X}. The
maximum error from the two basis states tested for the X gate was
$6.4 \times 10^{-6}$. The error induced in this operation is less
than a threshold of $1 \times 10^{-4}$ required for fault tolerant
quantum computing.

\begin{table}
\begin{tabular}{|c|c|c|}
\hline
State & Systematic Error & Typical Error \\
\hline
$\ket{0}$ & $2.3 \times 10^{-6}$ & $3.8 \times 10^{-6}$ \\
$\ket{1}$ & $4.9 \times 10^{-6}$ & $6.4 \times 10^{-6}$ \\
\hline
Maximum & $4.9 \times 10^{-6}$ & $6.4 \times 10^{-6}$ \\
\hline
\end{tabular}
\caption{Summary of X gate error.} \label{tab:X}
\end{table}

\section{Two Qubit Gates} \label{sec:TwoQubit}

\subsection{The CNOT Gate} \label{ssec:CNOT}

The CNOT gate is specified by
\begin{equation}
\Gamma_1X = \left[
	\begin{array}{cccc}
	1 & 0 & 0 & 0 \\
	0 & 1 & 0 & 0 \\
	0 & 0 & 0 & 1 \\
	0 & 0 & 1 & 0
	\end{array}
	\right],
\end{equation}
which may be created using the steps specified in Ref. \cite{HG03}. In
the following discussion of two qubit gates, if not explicitly stated,
all initial electron spin states are assumed to be
$\ket{\downarrow\downarrow}$.

\begin{figure*}
\subfigure[$\ket{00}$ initial state] 
	{ \includegraphics[width=8cm]{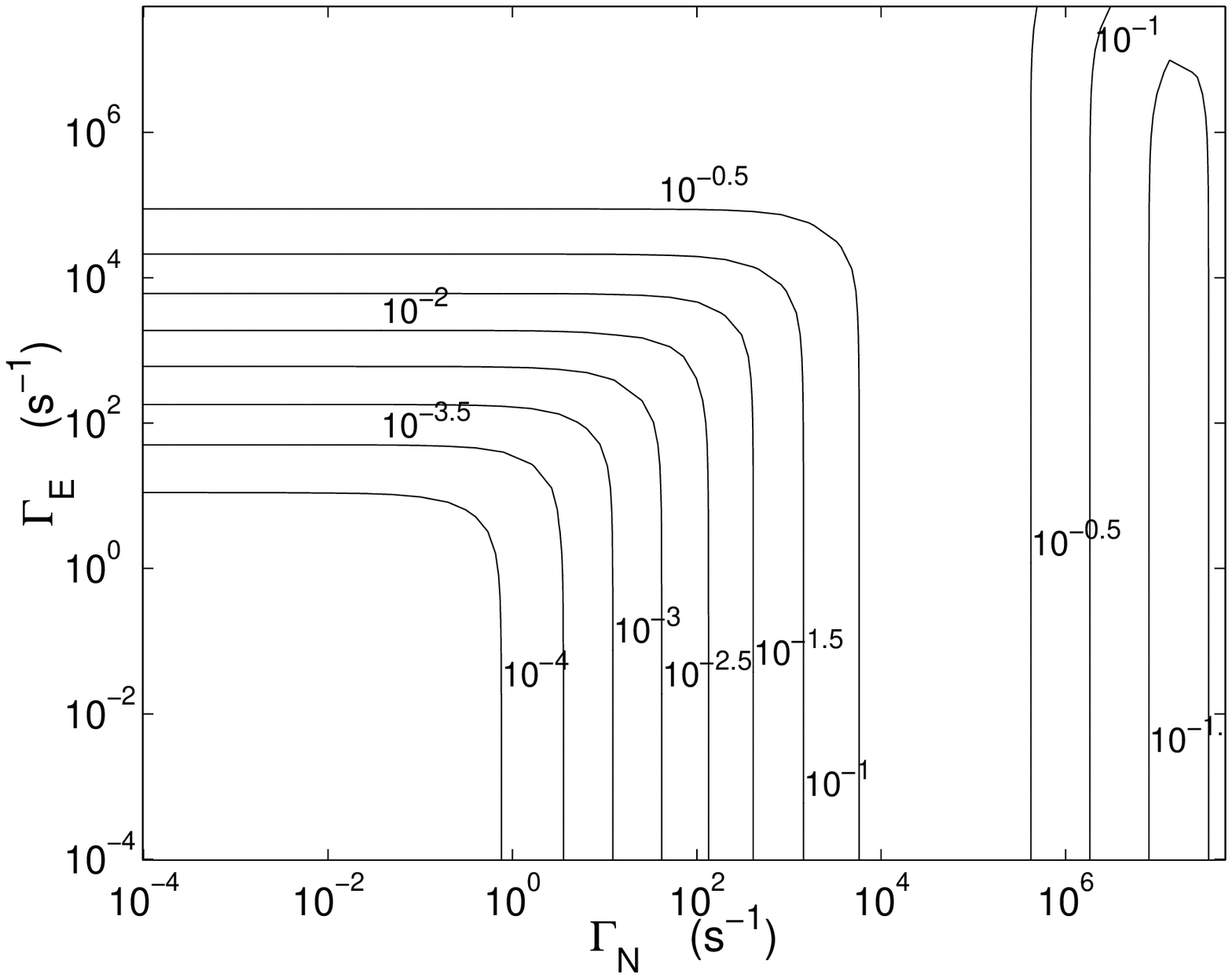} }
\subfigure[$\ket{01}$ initial state] 
	{ \includegraphics[width=8cm]{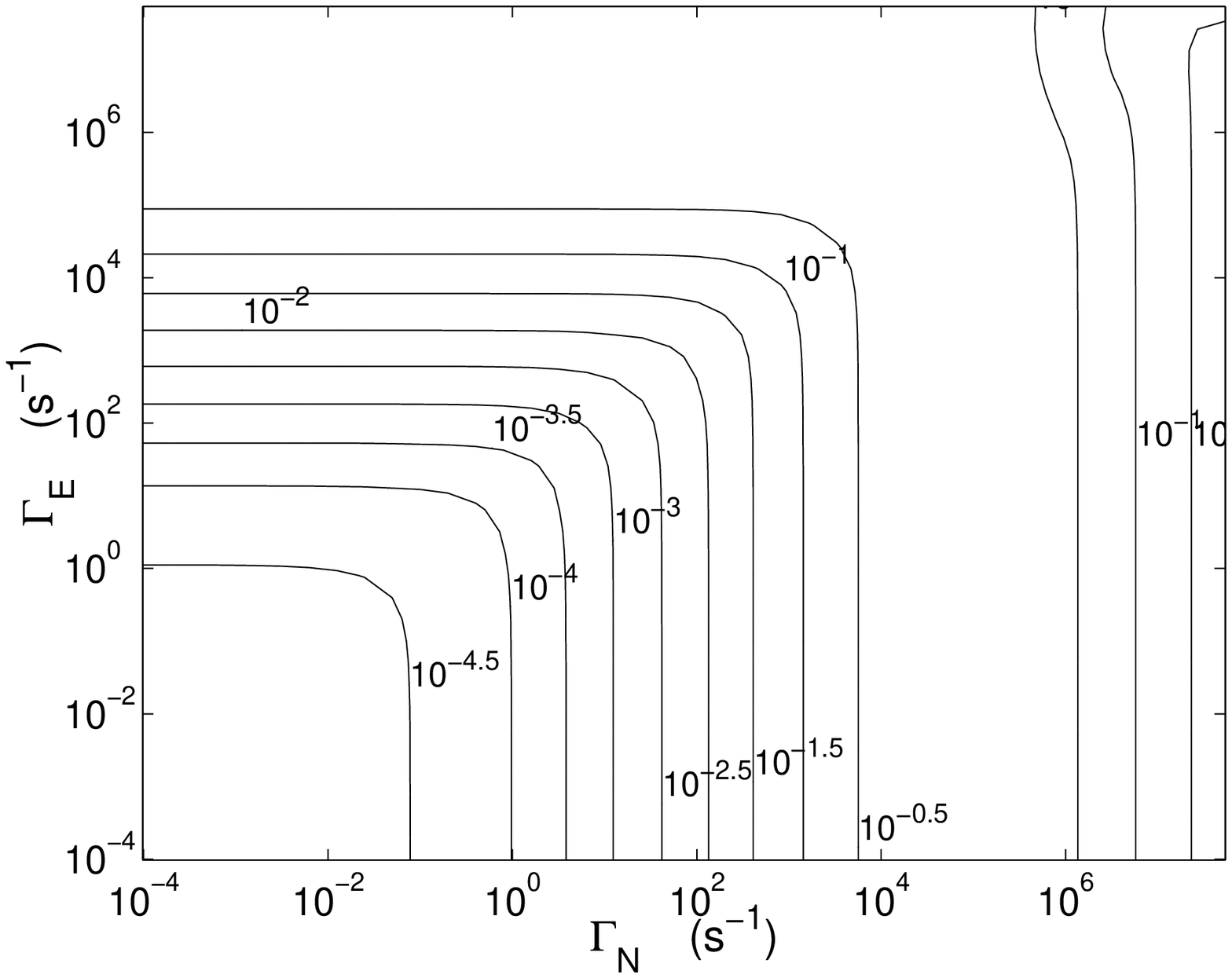} }
\subfigure[$\ket{10}$ initial state] 
	{ \includegraphics[width=8cm]{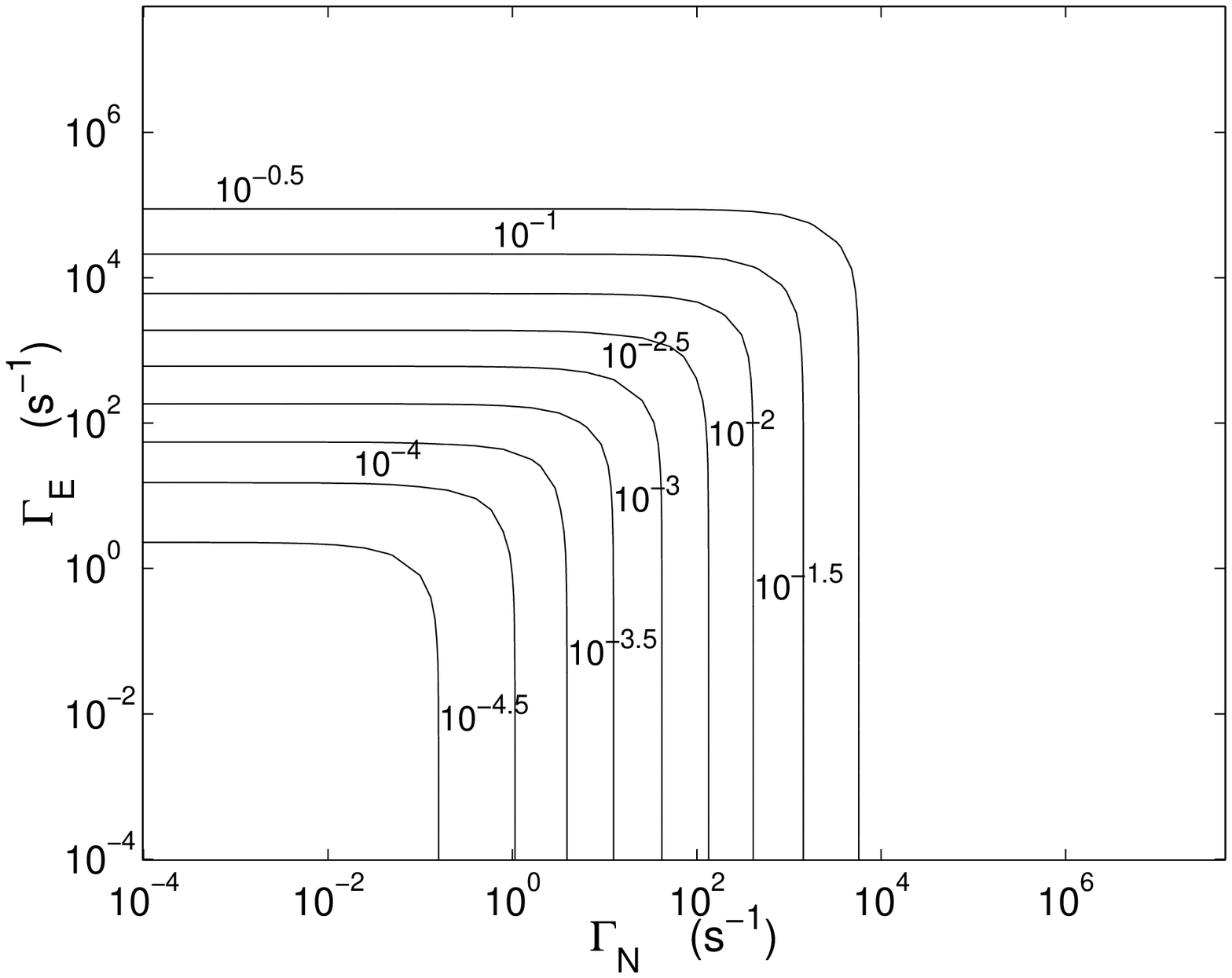} }
\subfigure[$\ket{11}$ initial state] 
	{ \includegraphics[width=8cm]{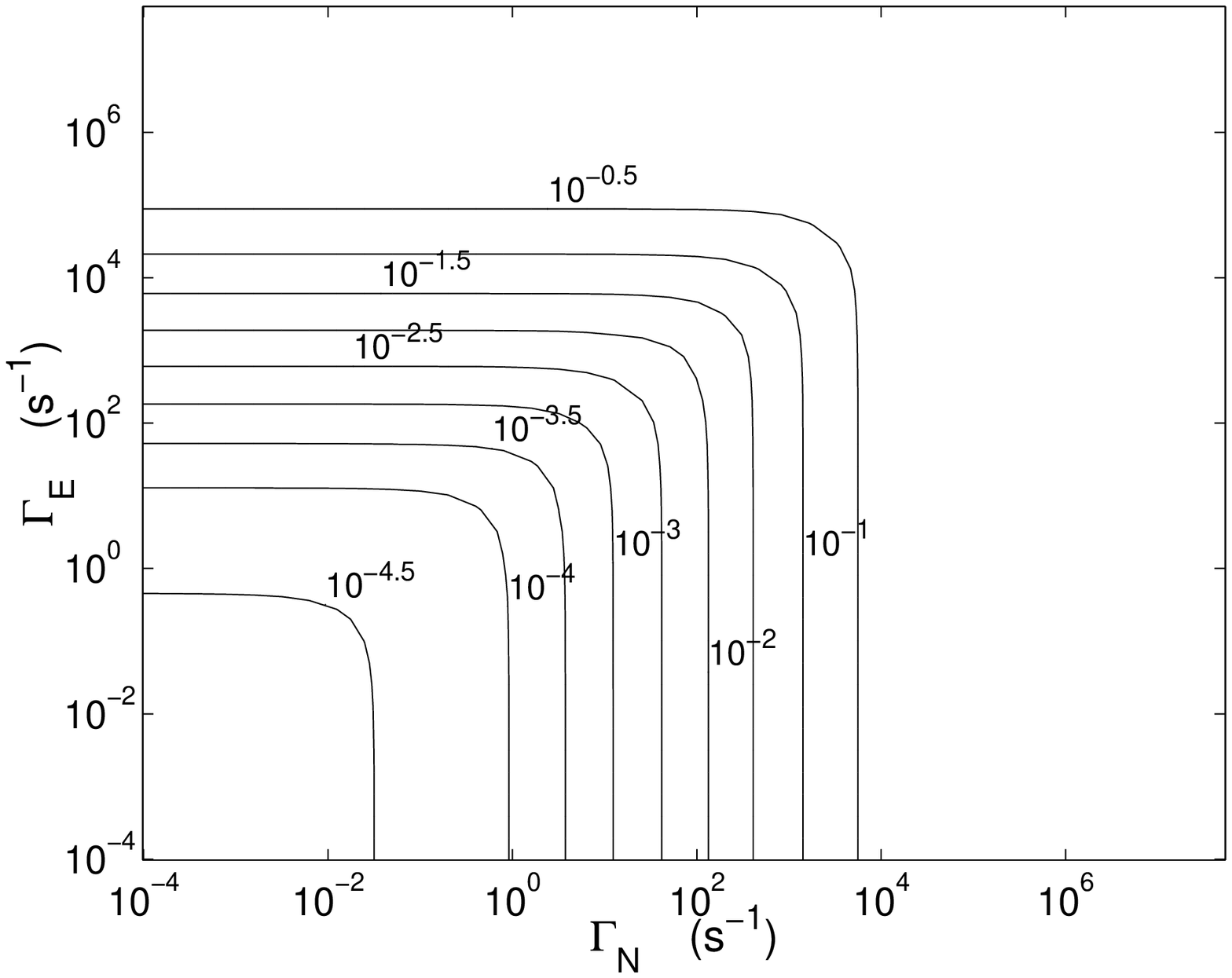} }
\caption{Contour plots showing error in the CNOT operation for
different rates of electronic and nuclear dephasing. Contour lines
connect, and are labelled by, points of equal error.}
\label{fig:CNOTFidelity}
\end{figure*}

The error in the CNOT in the presence of dephasing was found by
numerically solving the master equation [Eqs. (\ref{eqn:Master}) and
(\ref{eqn:L})] for the appropriate pulse sequence \cite{HG03}. For
different rates of dephasing, different fidelities are
obtained. Fidelities were calculated for each of the four
computational basis states, and for the evolution leading to the four
Bell states. Fidelities for the four computational basis states are
shown in Fig. \ref{fig:CNOTFidelity}.

Each of the states shows a minimum error when the rates of electronic
and nuclear dephasing are slow. The remaining error is due to
systematic error in the gate. Some sources of systematic error for the
CNOT operation include off-resonant effects, excitation into higher
electronic energy levels, and imperfections in the pulse sequences
(such as the break-down of the second order approximations used to
derive appropriate times for pulse sequence \cite{HG03}). Systematic
error for each of the states $\ket{00}$, $\ket{01}$, $\ket{10}$ and
$\ket{11}$ are $4.0 \times 10^{-5}$, $2.6 \times 10^{-5}$, $1.9 \times
10^{-5}$, and $2.9 \times 10^{-5}$ respectively. For evolution
starting in an initial Bell state, we find the systematc error is $3.5
\times 10^{-5}$, $3.4 \times 10^{-5}$, $1.9 \times 10^{-5}$ and $2.6
\times 10^{-5}$. Sytematic error for states resulting in a Bell state
are shown in Table \ref{tab:CNOT}.

In each of the four states, as the dephasing rate increases, the
fidelity decreases. In the limit of large dephasing rates, the
computational basis states tend to stay in their original states. This
is particularly evident from the graphs of $\ket{00}$ and $\ket{01}$
which have higher fidelity (lower error) at high dephasing rates. In
contrast, the states $\ket{10}$ and $\ket{11}$ have lower fidelities
(highter error) at high rates of dephasing. For example, the
$\ket{00}$ state stays in the $\ket{00}$ state after the CNOT
operation. At such high rates of dephasing, not even the single qubit
rotations described in Section \ref{ssec:XRot} apply, which form part
of the CNOT gate operation. For these states quantum coherence has
been lost. When we consider, for example, the state $\ket{\psi} =
1/\sqrt{2}(\ket{00} + \ket{01})$ we find that at a high dephasing rate
of $\Gamma_n = \Gamma_e = 54 \times 10^6 s^{-1}$ that the error of the
gate is $0.5$. In this case, quantum coherence has been lost between
the two states, and the qubit evolves to a completely mixed state.

Electronic dephasing rates play a much bigger role in two qubit gates
than in single qubit gates.

For the typical dephasing times, $T_{2_e}$ and $T_{2_n}$ [given in
Eq. (\ref{eqn:typicalDephasing1}), Eq. (\ref{eqn:typicalDephasing2})],
we find the error for the states $\ket{00}$, $\ket{01}$, $\ket{10}$
and $\ket{11}$ are $8.3 \times 10^{-5}$, $6.8 \times 10^{-5}$, $6.2
\times 10^{-5}$, and $7.2 \times 10^{-5}$ respectively. The error for
an initial state of one of the four Bell states is found to be
$6.0\times 10^{-5}$, $6.0 \times 10^{-5}$, $4.4 \times 10^{-5}$, and
$5.1 \times 10^{-5}$. Typical errors for states resulting in a Bell
state are shown in Table \ref{tab:CNOT}. This implies that under our
very simple decoherence model, the maximum error in the CNOT gate in
any basis state is $8.3 \times 10^{-5}$. This is only marginally under
the threshold of $1 \times 10^{-4}$ required for fault tolerant
quantum computation.

\begin{figure}
\begin{center}
\includegraphics[width=8cm]{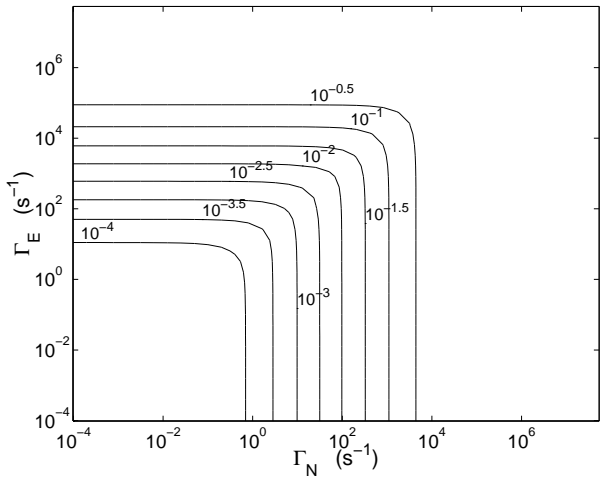}
\caption{Contour plot of the maximum error in basis and Bell output
states of the CNOT gate shown as a function of electronic and nuclear
dephasing rates. Contour lines connect, and are labelled by, points of
equal error.}
\label{fig:CNOTMinFidelity}
\end{center}
\end{figure}

Errors for each of the computational basis states are shown in Table
\ref{tab:CNOT}, and the maximum error of any of the four computational
basis states or states with evolution leading to a Bell state is
plotted in Fig. \ref{fig:CNOTMinFidelity}.

\begin{table}
\begin{tabular}{|c|c|c|}
\hline
State & Systematic Error & Typical Error \\
\hline
$\ket{00}$ & $4.0 \times 10^{-5}$ & $8.3 \times 10^{-5}$ \\
$\ket{01}$ & $2.6 \times 10^{-5}$ & $6.8 \times 10^{-5}$ \\
$\ket{10}$ & $1.9 \times 10^{-5}$ & $6.2 \times 10^{-5}$ \\
$\ket{11}$ & $2.9 \times 10^{-5}$ & $7.2 \times 10^{-5}$ \\
\hline
$\ket{00}+\ket{11}$ & $2.9 \times 10^{-5}$ &$7.0 \times 10^{-5}$\\
$\ket{00}-\ket{11}$ & $3.2 \times 10^{-5}$ &$7.3 \times 10^{-5}$\\
$\ket{01}+\ket{10}$ & $3.1 \times 10^{-5}$ &$7.2 \times 10^{-5}$\\
$\ket{01}-\ket{10}$ & $2.3 \times 10^{-5}$ &$6.4 \times 10^{-5}$\\
\hline
Maximum & $4.0 \times 10^{-5}$ & $8.3 \times 10^{-5}$ \\
\hline
\end{tabular}
\caption{Summary of CNOT gate fidelities.} \label{tab:CNOT}
\end{table}

These results are directly analogous to calculations made for the
adiabatic CNOT gate \cite{FWH03}. We have used the same noise model as
was used in their numerical simulations. In calculations for the
adiabatic CNOT gate at decoherence times, for electronic and nuclear
dephasing rates of $T_{2_e}=200 \mu s$ and $T_{2_n}=10s$ respectively,
the maximum error of any of the four basis states for the adiabatic
CNOT gate was found to be `just over $10^{-3}$' \cite{FWH03}. In
comparison, for the same conditions, we find that the maximum error in
the non-adiabatic gate is $7 \times 10^{-3}$. Under these conditions
both non-adiabatic and adiabatic gates give similar fidelities.

Where does this error come from at the dephasing rate specified above,
for the nonadiabatic CNOT gate? The error in an X rotation under these
conditions is $3 \times 10^{-6}$ and the error in the entangling
operation of the gate $U_m\left(\frac{\pi}{4}\right)$, is $4 \times
10^{-3}$ and therefore it is clear that the two qubit entangling
operation is the major source of error. When electronic decoherence
times are short, any electron mediated operations will be affected by
this decoherence. By increasing the exchange interaction strength, the
time required for the electron mediated operation may be reduced. For
example, at a strength of $J=0.0529 \mathrm{meV}$ the error decreases
to $1 \times 10^{-3}$ and the maximum error in the CNOT gate is also
$1 \times 10^{-3}$ for any of the computational basis states.

The advantage of nonadiabatic gates over adiabatic gates is that the
pulse schemes required are much simpler, faster, and as at the
conditions considered in Ref. \cite{FWH03}, the two schemes have
approximately the same fidelity. Considering the electronic
decoherence times measured in Ref. \cite{TLA+03} it is likely that
electronic dephasing rates are not as large as considered in Ref.
\cite{FWH03}. At lower rates of dephasing, we approach the systematic
error, which may be smaller for nonadiabatic gates than adiabatic
gates \cite{HG03}. Another distinct advantage of the nonadiabatic
gates is that \emph{any} two qubit gate may be made. This allows us to
construct two qubit gates directly (such as the swap gate), which are
faster and higher fidelity than expressing them as combinations of
CNOT gates and single qubit rotations. If the Kane computer was being
run from a digital clock cycle, non-adiabatic two qubit gates could be
controlled at discrete times, and do not require the continuous and
sophisticated pulse shapes required for adiabatic gates operating at
this speed and fidelity.

\subsection{The Swap Gate and Controlled Z Gate} \label{ssec:Swap}

Similar calculations to those calculated for the CNOT gate were
carried out for the swap gate and the controlled Z gate. The swap gate
is specified by the matrix
\begin{equation}
U_{Swap} = \left[
	\begin{array}{cccc}
	1 & 0 & 0 & 0 \\
	0 & 0 & 1 & 0 \\
	0 & 1 & 0 & 0 \\
	0 & 0 & 0 & 1
	\end{array}
	\right].
\end{equation}
The controlled Z gate is specified by the matrix
\begin{equation}
\Gamma_1Z = \left[
	\begin{array}{cccc}
	1 & 0 & 0 & 0 \\
	0 & 1 & 0 & 0 \\
	0 & 0 & 1 & 0 \\
	0 & 0 & 0 & -1
	\end{array}
	\right].
\end{equation}
The circuit which may be used to create the swap gate may be found in
Ref. \cite{HG03}.

\begin{figure}
\begin{center}
\includegraphics[width=8cm]{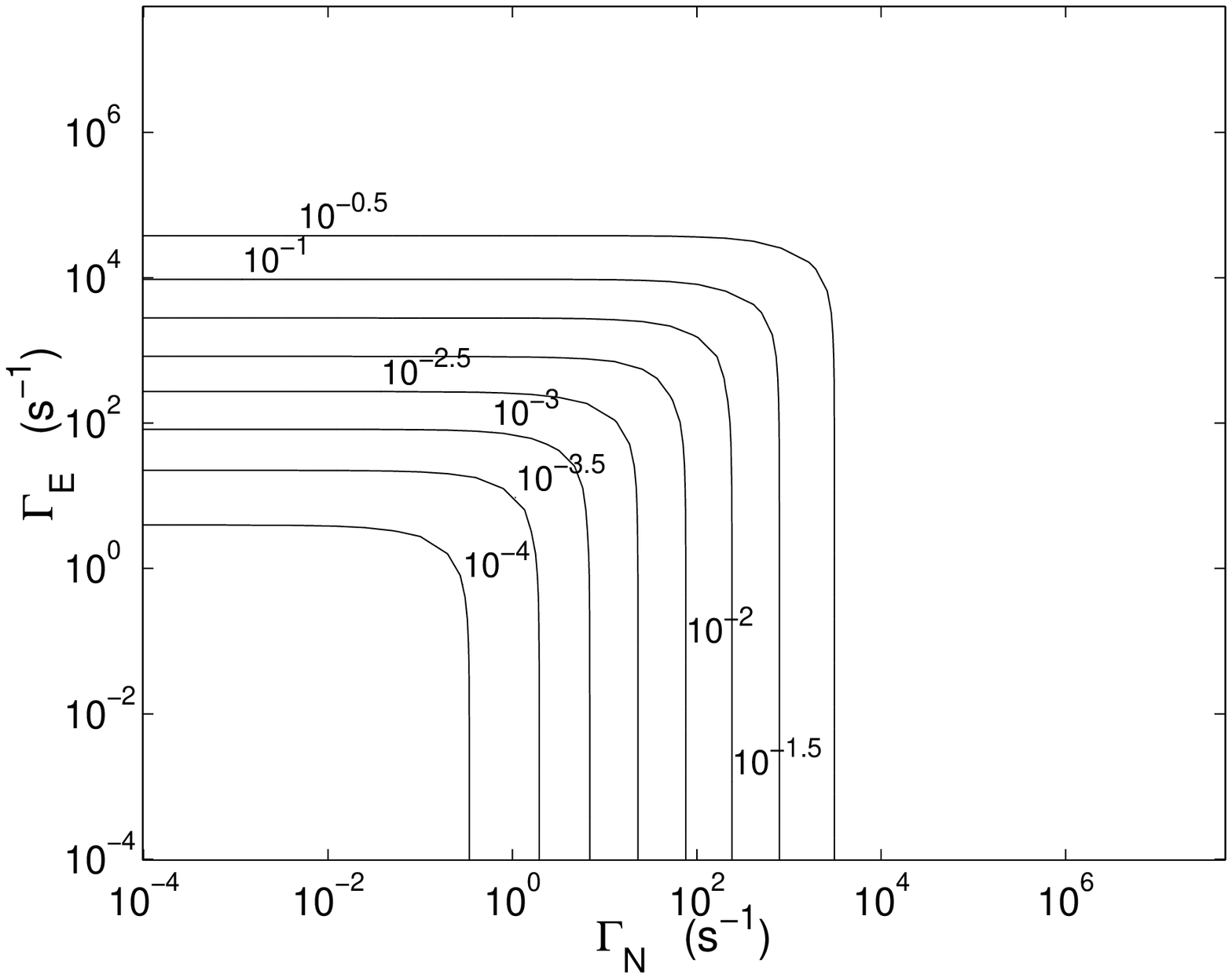}
\caption{Contour plot of the maximum error in basis states and Bell
output states of the swap gate shown as a function of electronic and
nuclear dephasing rates. Contour lines connect, and are labelled by,
points of equal error.}
\label{fig:SwapMinFidelity}
\end{center}
\end{figure}

\begin{figure}
\begin{center}
\includegraphics[width=8cm]{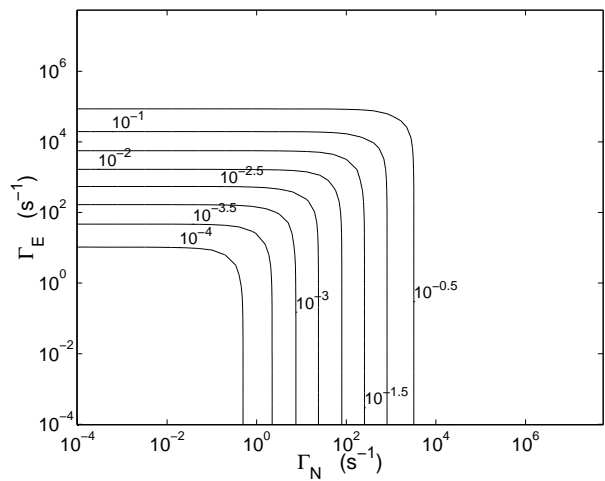}
\caption{Contour plot of the maximum error in basis states and Bell
output states of the $\Gamma_1{Z}$ gate shown as a function of
electronic and nuclear dephasing rates. Contour lines
connect, and are labelled by, points of equal error.}
\label{fig:CZMinFidelity}
\end{center}
\end{figure}

The master equation was solved numerically for each pulse
sequence. The error in the swap gate was calculated for each basis
state, and each state whose output state is a Bell state. Note that
for these two gates the states which give Bell states as output are
themselves Bell states. A separate simulation was completed for each
combination of nuclear and electronic dephasing times. The maximum
error of any of the basis states has been plotted in
Fig. \ref{fig:SwapMinFidelity} for the swap gate, and
Fig. \ref{fig:CZMinFidelity} for the control Z gate.

Similar features that were evident for the CNOT gate are visible in
these figures. The corresponding errors are shown in Table
\ref{tab:Swap} for the swap gate, and Table \ref{tab:CZ} for the
control Z gate.

\begin{table}
\begin{tabular}{|c|c|c|}
\hline
State & Systematic Error & Typical Error \\
\hline
$\ket{00}$ & $3.9 \times 10^{-5}$ & $9.0 \times 10^{-5}$ \\
$\ket{01}$ & $1.4 \times 10^{-5}$ & $7.9 \times 10^{-5}$ \\
$\ket{10}$ & $1.6 \times 10^{-5}$ & $8.0 \times 10^{-5}$ \\
$\ket{11}$ & $3.8 \times 10^{-5}$ & $8.9 \times 10^{-5}$ \\
\hline
$\ket{00}+\ket{11}$ & $5.3 \times 10^{-5}$ & $1.4 \times 10^{-4}$\\
$\ket{00}-\ket{11}$ & $7.4 \times 10^{-5}$ & $1.6 \times 10^{-4}$\\
$\ket{01}+\ket{10}$ & $1.7 \times 10^{-5}$ & $1.0 \times 10^{-4}$\\
$\ket{01}-\ket{10}$ & $1.5 \times 10^{-5}$ & $1.0 \times 10^{-4}$\\
\hline
Maximum & $7.4 \times 10^{-5}$ &  $1.6 \times 10^{-4}$ \\
\hline
\end{tabular}
\caption{Summary of swap gate error.} \label{tab:Swap}
\end{table}

\begin{table}
\begin{tabular}{|c|c|c|}
\hline
State & Systematic Error & Typical Error \\
\hline
$\ket{00}$ & $3.0 \times 10^{-5}$ & $8.1 \times 10^{-5}$ \\
$\ket{01}$ & $1.5 \times 10^{-5}$ & $6.6 \times 10^{-5}$ \\
$\ket{10}$ & $1.1 \times 10^{-5}$ & $6.2 \times 10^{-5}$ \\
$\ket{11}$ & $3.8 \times 10^{-5}$ & $8.9 \times 10^{-4}$ \\
\hline
$\ket{00}+\ket{11}$ & $3.6 \times 10^{-5}$ & $9.4 \times 10^{-5}$\\
$\ket{00}-\ket{11}$ & $3.3 \times 10^{-5}$ & $9.2 \times 10^{-5}$\\
$\ket{01}+\ket{10}$ & $1.7 \times 10^{-5}$ & $7.5 \times 10^{-5}$\\
$\ket{01}-\ket{10}$ & $1.1 \times 10^{-5}$ & $7.0 \times 10^{-5}$\\
\hline
Maximum & $3.8 \times 10^{-5}$ &  $9.4 \times 10^{-5}$ \\
\hline
\end{tabular}
\caption{Summary of controlled Z gate error.} \label{tab:CZ}
\end{table}

\section{Conclusion} \label{sec:Conclusion}

In conclusion, we have investigated the effect of dephasing on the
Kane quantum computing architecture. We used a simple model of
decoherence and investigated how this model affected proposed gate
schemes on the Kane quantum computer. For typical decoherence rates
[given in Eq. (\ref{eqn:typicalDephasing1}),
Eq. (\ref{eqn:typicalDephasing2})], these results are summarized in
Table \ref{tab:Fidelities}.

Each of the errors, for typical rates of dephasing, found here are
close to the error threshold required for fault tolerant quantum
computation. If the temperature is lowered, and coupling between
qubits is not considered a decoherence process as in
Ref. \cite{TLA+03}, it is likely that the typical decoherence times
for the Kane architecture may be further reduced, and therefore
unambiguously under the threshold required for fault tolerant quantum
computation.

Construction and operation of the Kane quantum computer is extremely
challenging. In the actual physical system there will undoubtedly be
noise and decoherence processes not considered in our simple physical
model. This substantial effort would never be able to achieve its
ultimate goal of a working quantum computer if there were fundamental
reasons why such a computer could not operate. In this paper we
investigated the one such effect on proposed gates for the Kane
quantum computer.  Our simulations indicate that errors due to
dephasing, the dominant form of decoherence in the Kane architecture,
do not rule out fault tolerant quantum computation.

\begin{table}
\begin{tabular}{|c|c|c|c|}
\hline \textbf{Gate} & \textbf{Typical Error} & \textbf{Systematic
Error} & \textbf{Time}\\
\hline
Z & $3.8 \times 10^{-6}$ & $3.8 \times 10^{-6}$ & $0.02 \mu s$\\
X & $6.4 \times 10^{-6}$ & $4.9 \times 10^{-6}$ & $6.4 \mu s$ \\ 
CNOT & $8.3 \times 10^{-5}$ & $4.0 \times 10^{-5}$ & $16.0 \mu s$\\ 
Swap & $1.6 \times 10^{-4}$ &  $7.4 \times 10^{-5}$ & $19.2 \mu s$\\
$\Gamma_1Z$ & $9.4 \times 10^{-5}$ & $3.8 \times 10^{-5}$ & $16.1 \mu s$ \\ 
\hline 
\end{tabular}
\caption{Times and errors for each of the gates investigated.}
\label{tab:Fidelities}
\end{table}

\acknowledgments The authors would like to thank G. J. Milburn for
support. This work was supported by the Australian Research Council,
the Australian government and by the US National Security Agency
(NSA), Advanced Research and Development Activity (ARDA) and the Army
Research Office (ARO) under contract number
DAAD19-01-1-0653. H.-S.G. would like to acknowledge financial support
from Hewlett-Packard.

\bibliography{gatesNoise2}

\end{document}